\documentclass[conference,compsoc]{IEEEtran}
\IEEEoverridecommandlockouts
\usepackage{cite}
\usepackage{amsmath,amssymb,amsfonts}
\usepackage{algorithmic}
\usepackage{graphicx}
\usepackage{textcomp}
\usepackage{makecell}
\usepackage{xcolor}
\usepackage{longtable}
\usepackage{fancyvrb}
\usepackage{listings}
\usepackage{alltt}
\usepackage{multicol}
\def\BibTeX{{\rm B\kern-.05em{\sc i\kern-.025em b}\kern-.08em
    T\kern-.1667em\lower.7ex\hbox{E}\kern-.125emX}}
\usepackage[absolute]{textpos}
\begin{document}

\bstctlcite{BSTcontrol}
\newcommand{\comment}[1]{}
\newcommand{\name}{AuthRank}

\begin{textblock}{17}(2,1.5)
\centering \noindent \large \emph{In Proceedings of the IEEE ACSAC Workshop on AI for Cyber Threat Intelligence (WAITI) 2026}
\end{textblock}

\title{\name: Scoring Source Credibility to \\Counter Poisoning in CTI LLM Pipelines}

\author{\IEEEauthorblockN{Austin Jia\IEEEauthorrefmark{1}, Avaneesh Ramesh\IEEEauthorrefmark{1}, Zain Shamsi\IEEEauthorrefmark{2}, Daniel Zhang, and Alex Liu}
\IEEEauthorblockA{\textit{Applied Research Laboratories, University of Texas at Austin}\\
\textit{Austin, Texas, USA} \\
}
\thanks{\IEEEauthorrefmark{1}Equal Contribution}
\thanks{\IEEEauthorrefmark{2}Corresponding Author. Email: zain.shamsi at utexas.edu}
}

\maketitle

\begin{abstract}
Large Language Model (LLM) pipelines have become essential for processing Cyber Threat Intelligence (CTI) to quickly extract updated threat information from open source feeds and reports for real time processing. However, these pipelines are susceptible to poisoning attacks, and previously proposed defenses can fail for CTI contexts as cyber threat information is often completely novel, and attackers can easily mimic the stylistic conventions. We propose that the robustness of LLM pipelines can be improved by applying source credibility algorithms on document corpora, using PageRank as an example. We introduce \name, a credibility score computed from citation networks built using explicit citations, LLM-inferred citations, and claim-level entailment, enhanced with time decay and author credibility. Our experiments on the standardized MS MARCO dataset demonstrate improved accuracy under poisoning, and we show proof-of-concept performance on CTI documents and feeds.
\end{abstract}


\section{Introduction}

Large Language Models (LLMs) have grown in use immensely due to their extraordinary generative capabilities and diverse applications. The integration of LLMs into cybersecurity operations \cite{cuong_nguyen_towards_2025, hanks_recognizing_2022, alam_ctibench_2024, schwartz_llmcloudhunter_2025, shah_mad-cti_2025, buchel_sok_2025} marks a paradigm shift, where unstructured Cyber Threat Intelligence (CTI) reports can be ingested and analyzed, and actionable Indicators of Compromise (IoCs) can be extracted and fed into downstream systems. However, the ever-evolving nature of cybersecurity and the huge volumes of data available from multiple threat feeds \cite{AlienVault, threatfeeds, Malwarebytes, tweetfeed} means that a large amount of contextual information relevant to this processing is constantly being generated. This highlights the inherent limitation of LLMs: their lack of direct access to current domain-specific knowledge, as well as their propensity to hallucinate in contexts where actual data is not available as a point of reference.

To address this issue, LLM pipelines that ground generation in external document corpora have become the dominant approach for operationalizing LLM usage in CTI systems \cite{lewis_retrieval-augmented_2020, buchel_sok_2025, fayyazi_advancing_2024, tellache_advancing_2025}. These range from single-shot Retrieval-Augmented Generation (RAG), which retrieves relevant passages in one pass, to agentic pipelines where LLM-based agents autonomously navigate documents, follow citation chains, and synthesize threat intelligence across multiple reasoning steps \cite{hamzic2026beyond, yao2022react}. Both architectures allow LLMs to access an up-to-date knowledge base, providing the necessary context to complete tasks such as summarizing threat actor profiles, linking IoCs, or explaining complex attack techniques.

However, this dependence on external corpora introduces a new attack vector for infusing misinformation into CTI pipelines. Both RAG and agentic pipelines have been shown to be susceptible to poisoning attacks \cite{chen_agentpoison_2024, zou_poisonedrag_2025, roychowdhury_confusedpilot_2024}. In RAG, attackers craft malicious injections optimized for retrieval inclusion. In agentic pipelines, the threat is compounded: an agent that iteratively retrieves and reasons over documents may follow a chain of poisoned sources across multiple steps, or a poisoned document may steer the agent's trajectory toward malicious conclusions. \comment{To go along with this, the assumptions that the attacker must have at least partial access to the database are magnified in our context, as open source CTI feeds are essentially a shared data source that the attacker has full access to and can use to tune their malicious injections.} With complete access to open source CTI feeds, adversaries are easily able to inject deceptive elements such as falsified threat reports, manipulated IoCs, misleading vulnerability data, or poisoned mitigation advice. Since CTI is a shared resource between many organizations, this corrupts the entire security pipeline by amplifying this misinformation. The consequences range from wasted resources by security analysts who may chase phantom threats and deploy ineffective mitigations, to catastrophic misconfigurations and the obscuring of genuine, ongoing attacks \cite{shafee_false_2025}. Indeed, if the attacker's goal is to disrupt operations and cause havoc by misconfiguration of key services, carefully crafted fake CTI could well accomplish that goal by itself. Consequently, verifying CTI authenticity within LLM pipelines represents a fundamental operational requirement.

\subsection{Motivation}
Prior defenses to corpus poisoning have put forward various techniques, such as using modified majority voting schemes, having multiple LLMs verify relevance, adding redundant safe data to the retrieval, clustering retrieved texts in embedding space, and involving humans in the loop \cite{vonderhaar_surveying_2025}. These techniques mainly focus on analyzing the retrieved subject matter. However, these defenses can fail for CTI contexts as cyber threat information that is completely new is difficult to verify, and sophisticated threat actors can mimic legitimate CTI formats, terminology and stylistic conventions. Thus, we propose that as an alternative to interrogating what the information \emph{says}, we can also focus on where it originates and how it propagates.

This approach largely mirrors tactics used in search engine ranking. Given that corpus ingestion in LLM pipelines is fundamentally a search and reasoning problem, we believe that leveraging the decades of literature on search engine ranking to guard against irrelevant or misleading information is a natural fit. We turn to one of the most well-known ranking algorithms in PageRank \cite{page1999pagerank}, which was used by Google for ranking webpages based on inbound hyperlinks. \comment{All pages are initialized with ranks equal to $1/N$ where $N$ is the total number of pages, after which the ranks are recursively redistributed to their outward hyperlinks until the system stabilizes.} In essence, PageRank measures the prestige and amount of contextual support for each page from other sources. Documents in a corpus contain content-based links that are analogous to links between webpages. Therefore, we propose that the robustness of modern LLM pipelines can be improved by applying source credibility algorithms on corpora, using PageRank as an example, creating a signal which is orthogonal to content-based defenses and can be combined to improve defense against poisoning attacks.

\subsection{Contributions}
In our work, we introduce a PageRank-derived authority score that we call \emph{\name} as a credibility layer against corpus poisoning attacks. We first build a citation network from a document corpus using explicit citations, LLM-inferred citations, and claim-level entailment. We then incorporate refinements such as time decay to counter bias against recent documents, and author credibility to leverage author reputation. We describe how \name{} can be integrated into both single-shot RAG and multi-step agentic pipelines. In our experiments, we demonstrate that our algorithm improves accuracy of answers on a standardized dataset, and also show its performance on two CTI misinformation case examples.

\section{Related Work}

\noindent \textbf{Prior Defenses.}
Prior works have used various approaches to defend against corpus poisoning attacks and ensuring retrieved content is relevant to the query. In \cite{yoran_making_2024}, the authors used Natural Language Inference (NLI) techniques to make retrieval more robust, and show that training the model for the task can further improve performance. RobustRAG \cite{xiang_certifiably_2024} employs an isolate-then-aggregate strategy that separates each retrieved document and generates an LLM response from each before aggregating them for a final robust response. Similarly, other works have explored \emph{Knowledge Expansion} \cite{su_towards_2025, zou_poisonedrag_2025}, which expands the retrieval context in order to drown out the malicious injection. TrustRAG \cite{zhou_trustrag_2025} takes this a step further to identify malicious documents before the retrieval by clustering them in embedding space, in order to defeat attacks which generate a large amount of poisoned injections that might overcome the earlier strategies. The closest related work to our effort is ClaimTrust \cite{qian_claimtrust_2025}, which uses a modified PageRank-inspired algorithm to propagate trust scores across documents. However, ClaimTrust allows  multiple edges between nodes as well as negative weight edges. We found this would give false claims high authority if they are mixed in with many true claims, and negative weight edges could be abused to downrank legitimate documents.

\noindent \textbf{CTI Pipelines and Misinformation.}
Fayyazi et al. \cite{fayyazi_advancing_2024} showed that RAG performs better for CTI use cases than other methods, and Tellache et al. \cite{tellache_advancing_2025} presented the performance of a CTI RAG pipeline by retrieving relevant documents via CTI APIs from VirusTotal and CrowdStrike to correlate incidents with historical cases. Their results showed strong answer relevance and groundedness, and was validated by cybersecurity professionals. Conversely, Shafee et al. \cite{shafee_false_2025} demonstrated how adversarial text generation techniques can create fake cybersecurity information that misleads classifiers, degrades performance, and disrupts system functionality, leading to real-world consequences of wasted time and resources. Huang et al. \cite{huang_can_2025} evaluated how well generated CTI can be detected by humans as well as traditional Natural Language Processing (NLP) and machine learning techniques, concluding that humans cannot distinguish the generated content regardless of their background, and a multi-step approach is needed.

\begin{figure*}[ht]
    \centering
    \includegraphics[width=7in]{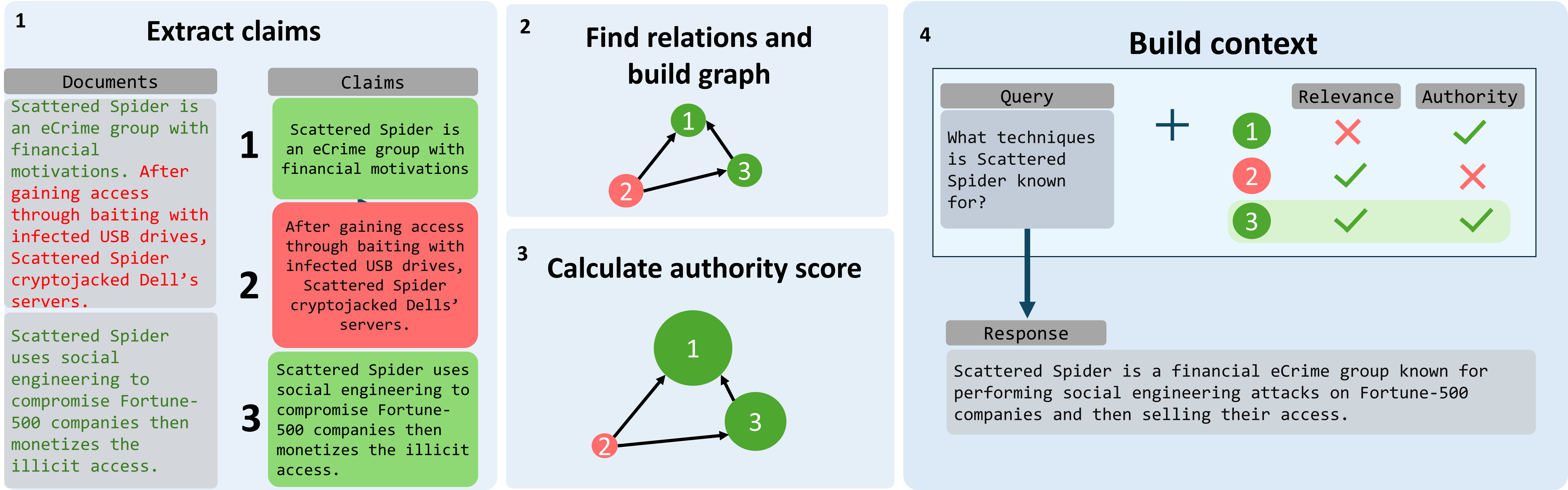}
    \caption{Overview of our system. Example shows graph building using claim extraction, where red = poisoned injection and green = legitimate information.}
    \label{fig:overview}
\end{figure*}

\section{Problem Definition}

An LLM pipeline takes a user query $q$ and a document corpus $\mathcal{D} = \{d_1, ..., d_n\}$, retrieves a subset of relevant documents, and generates an output $y = \texttt{LLM}(q \oplus \mathcal{C})$ where $\mathcal{C}$ is the accumulated retrieved context and $\oplus$ is concatenation. To formalize the retrieval step in a general manner for a user query $q$, the pipeline computes:
\begin{equation}
\label{eq:relevance}
    \mathcal{C} = \underset{d \in \mathcal{D}}{\text{top-}k}[\texttt{sim}(q,d)]
\end{equation}
where $\texttt{sim}(q,d)$ is a similarity function determined by the pipeline. In single-shot RAG, $\mathcal{C}$ is retrieved in one pass, and $\texttt{sim(q,d)}$ is usually cosine similarity $\cos(\phi(q), \phi(d)) \geq \theta$, where $\theta$ is a minimum threshold. In agentic pipelines, $\mathcal{C}$ is built dynamically over multiple steps as the agent retrieves, reads, and reasons over documents and similarity to the query is usually some form of attention over the accumulated and summarized context by the agent or subagents. Eventually, a set of top $k$ chunks of information are chosen to be included in $\mathcal{C}$ to produce $y$.

The adversary attempts to corrupt $\mathcal{D}$ by injecting malicious documents $\mathcal{D}_m = \{d_1^m, \dots, d_p^m\}$ targeting a specific query $q$ by solving:
\begin{equation}
    \max_{d^m \in \mathcal{D}_m} \ \mathbb{P}[ d^m \in \mathcal{C} ] \quad \text{s.t.} \quad \texttt{sim}(q, d^m) \geq \theta
\end{equation}
The attack succeeds when the context $\mathcal{C}$ for query $q$ contains a threshold number $M$ of malicious documents, such that the LLM bases its output $y$ on poisoned context:
\begin{align}
y &= \texttt{LLM}(q \oplus \mathcal{C}) \quad \text{where} \nonumber \\
\mathcal{C} &\subseteq \underset{d \in \mathcal{D} \cup \mathcal{D}_m}{\text{top-}k} [\texttt{sim}(q,d)]
\quad \text{and} \quad
|\mathcal{C} \cap \mathcal{D}_m| \geq M
\end{align}
In agentic pipelines, this threat is compounded: malicious documents encountered in early steps can steer subsequent retrievals, creating a chain of poisoned context across the agent's trajectory.

In our threat model for CTI, we might well have situations where there is little existing information in $\mathcal{D}$ for new emerging threats. This provides the attacker a favorable opportunity to inject malicious information to ensure it gets included in the retrieved context. We assume a non-adaptive adversary that optimizes injections for retrieval inclusion but is unaware of the authority layer.

\section{Methodology}

\subsection{Overview}
Given a document corpus, we explore three different ways to build a graph structure that represents the data set: explicit citations, inferred citations, and claims extracted from the documents. An overview of our approach using claim extraction is shown in Figure \ref{fig:overview}. Once we have a graph, we calculate the PageRank score over the graph which is then enhanced with additional measures, i.e., time decay to counter bias against recent documents which could provide important information about emerging threats, and author credibility to leverage author reputation. The resulting authority score, which we call \emph{\name}, is computed once as a preprocessing step and can be consumed by any pipeline architecture by instructing the LLM to incorporate the score when it determines document relevance. Next, we delve into the details of each step.

\subsection{Graph Construction}

We use three different techniques to build the corpus graph. We describe all three for the sake of completeness in this paper. However, due to space restrictions, we only present results using the ``inferred citations" method in the experiments.

\subsubsection{Explicit Citations}
The most direct construction treats each document as a vertex and adds a directed edge from a citing document to each work it references. While straightforward, two issues limit its practicality. First, enterprise CTI stores rarely carry explicit citation links between blog posts, vendor reports, and feed entries. Second, a citation does not imply agreement: a newer document may cite an older one precisely to refute it, yet the refuted document still accrues authority. Both motivate the inferred-citation approach below.

\subsubsection{Inferred Citations}
For this approach, we generate a similar graph as above, but use inferred citations, rather than explicit ones. We determine the commonality of the topics between two documents by using an LLM (we use Gemma 2-27B) to compare the contents. We prompt the LLM to consider claims and keywords, providing a specific metric of comparison (see Appendix \ref{sec:appendix} for full prompt). From this, the LLM determines the degree of commonality between the contents as a value between 0 and 1. This value is used as the weight of the connection between the two documents in the citation network.

We make our inferred citation directional by positing that if the contents of Document $B$ can be considered as a continuation of the discussion presented in Document $A$, Document $B$ cites Document $A$. The directionality of this citation is also determined by temporal factors, where Document $B$ can only cite Document $A$ if Document $B$ was published after Document $A$. If these conditions are met, then we add an edge from vertex $B$ to vertex $A$. Also, since LLM inference is expensive, we only run LLM inference for document pairs with a cosine similarity higher than $0.5$. We reason documents with a lower cosine similarity will likely have a negligible agreement score. This approach can take a few hours to complete initially, but can likely be sped up with newer models, and only updating subsets of the graph in a running pipeline.

This method addresses the issues mentioned above with the explicit citations approach. If newer documents provide conflicting information compared to older documents, the newer information will likely not result in an inferred citation link to the older information, regardless of an explicit citation link between the documents. This also makes our framework extensible to practical LLM pipelines that lack obvious graphical relationships. However, we do note some drawbacks. We noticed that incorrect connections can still be made between clean and malicious documents, largely because malicious documents still contained factually correct information. We also expect a larger model to do better than the Gemma 2 LLM at inferring citations between related documents.

\subsubsection{Claim Extraction}
We also explored a finer-grained variant in which vertices are individual claims rather than documents. We extract claims from each document using Gemma 2 and add an edge from claim A to claim B when a pretrained NLI model (RoBERTa-large-MNLI) judges that A entails B, treating entailment as a stricter form of the agreement relation used above. The resulting graph is far denser and yields good separation between clean and malicious PageRank scores, but the volume and granularity of claims made it difficult to map scores back to documents at inference time. Grouping or summarizing claims \cite{edge_local_2025} is a plausible remedy, and we leave this to future work.

\begin{table*}[ht]
\centering
\caption{\name{} Scoring Efficacy Against CTI Data Poisoning}
\label{tab:cti_one_results}
\begin{tabular}{|l|l|l|l|l|c|c|}
\hline
\textbf{Test APT} &
\textbf{Query Focus} &
\textbf{Correct Answer} &
\textbf{Poisoned Info} &
\thead{\textbf{Top Correct Source} \\ \textbf{(\name{} Score)}} &
\thead{\textbf{Top Poison} \\ \textbf{\name{} Score}} &
\textbf{Outcome} \\
\hline
\hline
CozyBear & Group affiliation & Russia & China & Wikipedia (0.92) & 0.31 & Correct \\
FancyBear & Organization nature & State-sponsored & Independent firm & CrowdStrike Blog (0.87) & 0.29 & Correct \\
LazarusGroup & Attack targets & Financial/crypto & Education/OSS & Wikipedia (0.85) & 0.33 & Correct \\
ScatteredSpider & Group purpose & Cybercriminal & Corporate red team & Conflicting sources* (0.22) & 0.22 & No Answer \\
ScatteredSpider & Term meaning & Cybercriminal group & Desert spider & Cyware Blog (0.94) & 0.25 & Correct \\
\hline
\end{tabular}
\par\vspace{0.5ex}
\hspace*{-0.70\linewidth}*This query only retrieved poisoned results
\end{table*}

\subsection{Calculating the Authority Score}
Since the attacker targets the similarity function $\texttt{sim}(q, d)$, we seek to augment this formula by adding an authority score $\alpha(d)$ that is considered along with the similarity result.

\subsubsection{PageRank}
Given a graph built using our techniques above, for each node $d_i$ we start by calculating its PageRank score \cite{page1999pagerank}:
\begin{equation}
    \alpha(d_i) = \frac{1 - \beta}{n} + \beta \sum_{d_j \in \mathcal{L}_{\text{in}}(d_i)} \frac{w_{ji}}{\sum_{d_k \in \mathcal{L}_{\text{out}}(d_j)}w_{jk}} \alpha(d_j)
\end{equation}
where $\beta$ is set to $0.85$, $\mathcal{L}_{\text{in}}(d_i)$ denotes inbound citations, $\mathcal{L}_{\text{out}}(d_j)$ denotes outbound citations, $n$ is the number of nodes and $w_{ji}$ is the weight of the edge from $d_j$ to $d_i$. We then enhance this score with additional post processing.

\subsubsection{Time-Decayed Rank}

Due to the importance of considering newer threat information, we implement a time decay such that the score of older documents is scaled by the difference from their creation date to the current date. This scaling occurs as a percentage of the document’s current authority score based on the distance from the current date.

Let $t(d_i)$ denote the age of document $d_i$ and $r$ be a relevance period after which a document's authority begins to decline. Define a linear time-decay factor:
\begin{equation}
\tau(d_i)=\left\{
\begin{array}{ll}
    1 & t(d_i) \leq r \\
    \max(0, 1 - \lambda \cdot (t(d_i) - r)) & t(d_i) > r \\
\end{array}
\right.
\end{equation}
where $\lambda$ is a decay rate hyperparameter. The time-decayed authority score is then computed as:
\begin{equation}
    \alpha'(d_i) = \tau(d_i) \cdot \alpha(d_i)
\end{equation}

In our testing, we set $t(d_i)$ to be the age in months and use $\lambda = 0.01$, but these parameters can be set to age out documents based on the required focus of the system and time spanned by the specific document corpus.

\subsubsection{Author Credibility}
We also consider author credibility in the scenario where a new document may deserve an increased initial authority due to its credible authors. 

Let $\mathcal{D}_x$ be the set of prior documents authored by author $x$. The credibility score for author $x$ is then defined as an average over their prior documents:
\begin{equation}
    C(x) = \frac{1}{|\mathcal{D}_x|} \sum_{d \in \mathcal{D}_x} \alpha'(d)
\end{equation}
The cumulative author credibility for document $d_i$ is averaged over all authors of that document:
\begin{equation}
    \gamma(d_i) = \frac{1}{|A(di)|}\sum_{x \in A(d_i)} C(x)
\end{equation}
where $A(d_i)$ is the set of all authors of document $d_i$. 

\subsubsection{Computing \name}
Armed with the above, we calculate our final score $R(d_i)$, which we call \name, as the min-max normalized sum of time-decayed rank and author credibility:
\begin{align}
\label{eq:authscore}
    R(d_i) &= \frac{s_i - \min_j s_j}{\max_j s_j - \min_j s_j} \quad \text{where} \nonumber \\
    s_i &= \alpha'(d_i) + \gamma(d_i)
\end{align}
and $\min_j$ and $\max_j$ are computed over all documents $d_j \in \mathcal{D}$. This yields a score in the range [0,1] for each document. The weighting of terms $\alpha'(d_i),\gamma(d_i)$ can be adjusted based on confidence in the author attribution of the dataset.

\subsection{Pipeline Integration}
In RAG pipelines, we employ a two-pass ranking strategy to balance relevance and authority.
In the first pass, we retrieve the top-$2k$ (instead of top-$k$) chunks most similar to the input prompt based on their cosine similarity as in \eqref{eq:relevance}. In the second pass, we re-rank these $2k$ candidates using their \name{} score. The top $k$ chunks from this re-ranked list are then selected as context for the language model. Thus, combining \eqref{eq:relevance} and \eqref{eq:authscore} we get:
\begin{align}
    \mathcal{C} &= \underset{d \in \mathcal{D}}{\text{top-}k} [ R(d) \cdot \mathbf{1}_{ \{ d \in \mathcal{C}_{2k} \}} ] \quad \text{where} \nonumber \\
    \quad \mathcal{C}_{2k} &= \underset{d \in \mathcal{D}}{\text{top-}2k} [ \texttt{sim}(q, d) ]
\end{align}
This two-stage approach addresses a key limitation of using a single weighted sum of cosine similarity and authority (e.g. $\omega \cdot R(d_i) + (1 - \omega) \cdot \texttt{sim}(q, d_i)$), where high-authority documents can overshadow more relevant but lower-authority ones. By first filtering for relevance and then re-ranking by authority, we ensure that only semantically related documents are considered, while still promoting trustworthy sources. The LLM is provided the re-ranked document list and \name{} for each document.

In agentic pipelines, the \name{} score is attached as metadata to each document and consulted at each reasoning step: the agent prioritizes high-authority documents for reading, weights claims by source authority during synthesis, and defers to higher-authority sources when resolving conflicts. Rather than a single re-ranking pass, the agent continuously applies authority scoring across its multi-step trajectory, which is particularly valuable in CTI contexts where an agent investigating an emerging threat may cross-reference many sources and a single poisoned document should not derail the investigation. In both cases, the system prompt is enhanced with rules to consider the \name{} score.

\section{Experiments and Results}
We performed preliminary experiments using a standard dataset and two conceptual cases with CTI misinformation. Since the \name{} score is computed offline and provided separately, our setup is agnostic to the type of pipeline.

\subsection{Standard Dataset}
We began our evaluation on poisoned versions of the MS MARCO \cite{bajaj_ms_2018} question answering dataset, as constructed by \cite{zou_poisonedrag_2025}. We present results ranging from one poisoned document to five poisoned documents for each question in the dataset, with five being the highest level of poisoning in \cite{zou_poisonedrag_2025}. After constructing the graph using the inferred citation method, we computed \name{} scores and then asked our LLM to answer all the questions in the dataset with instructions to prioritize high authority documents.

\begin{figure}[t]
    \centering
    \includegraphics[width=3.05in]{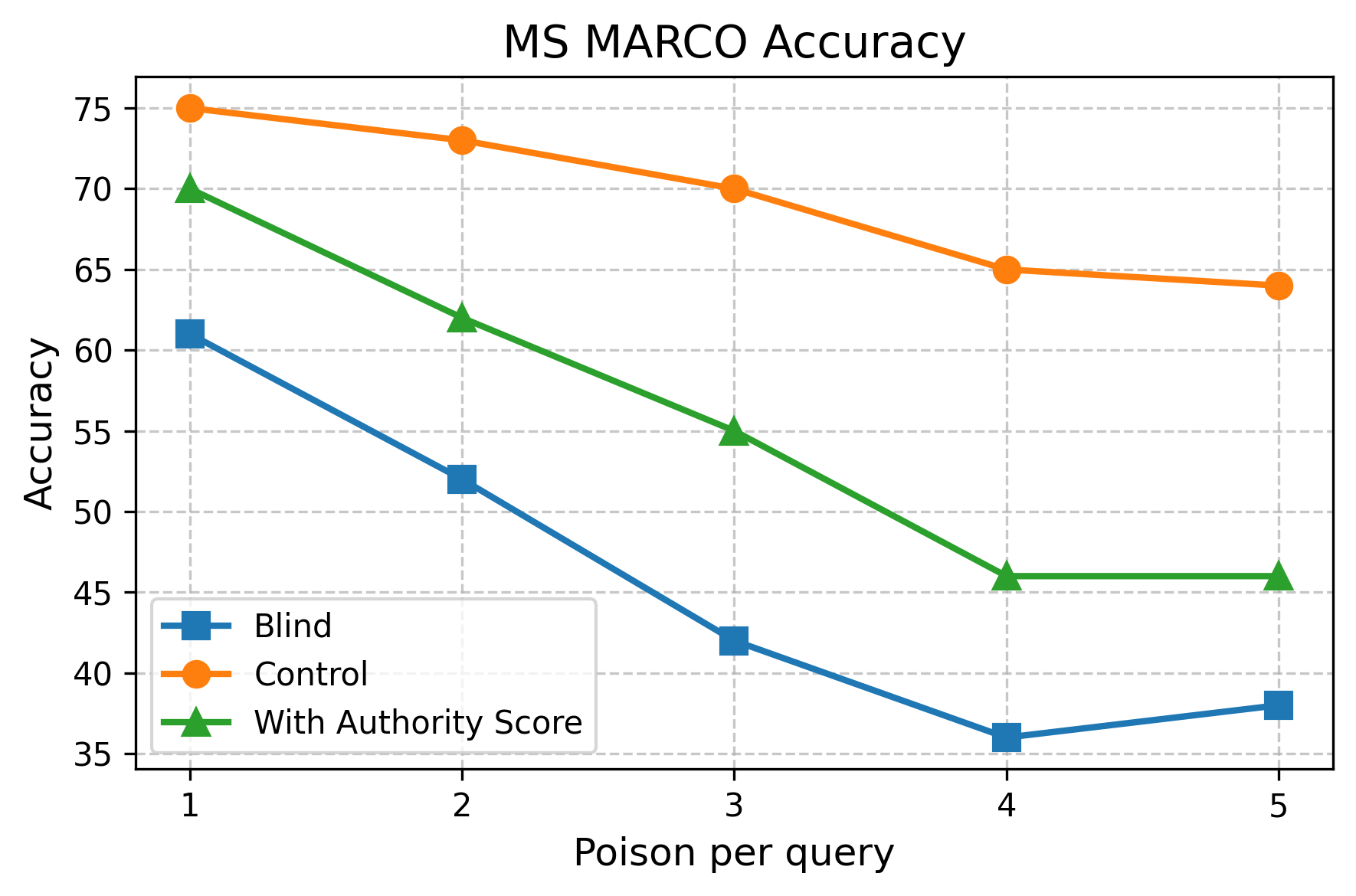}
    \caption{Accuracy of \name{} on poisoned MS MARCO dataset.}
    \label{fig:msmarco-accuracy}
\end{figure}

Our accuracy in this experiment is calculated based on whether the LLM found the correct answer in the corpus or whether it provided the poisoned answer. The results with increasing poisoning levels is shown in Figure \ref{fig:msmarco-accuracy}. Blind accuracy represents the performance without any poisoning protection, whereas the control represents the performance with all poisoning removed from each retrieval. Observe that \name{} is able to improve accuracy of the LLM by about 10-15\% in most cases, with difficulty increasing along the x-axis. We must note that this dataset does not include author and time metadata, hence we are unable to use the benefits those enhancements provide.

\subsection{CTI Case Studies}

In this section, we show two proof-of-concept CTI examples. Our corpus is structured to mimic a collection of articles and threat reports from cyber security organizations (e.g., CrowdStrike, Darktrace) and any other public news and data about threat actors (e.g., Wikipedia, Reuters) that may be ingested into a CTI database. After processing, we end up with about 300 excerpts used to build our graph.

In the first scenario, an attacker tries to drown out previously known information in a threat database using poisoned blog posts and articles. To test such a scenario, we create test cases that each contain a query about a known APT, and a list of ten poisoned chunks that attempt to trick the pipeline into retrieving them so their contents may manifest in the final LLM inference output. In total, we accumulate five test cases shown in Table \ref{tab:cti_one_results}, which lists the APT, the focus of the query, the correct answer, the incorrect answer encouraged by the attacker, the scores of the highest ranked retrieved documents (both actual and poison), and the outcome. For example, in the first row the query asks about the affiliation of the CozyBear APT, with the poisoned samples attempting to get the LLM to answer with China; however, the LLM correctly deferred to Wikipedia as the main source. Most cases were similar except for test four, where even though the LLM only retrieved poisoned documents in the first pass, it rightly refused to provide a conclusive answer due to none of the documents being authoritative.

\begin{figure}[t]
    \centering
    \includegraphics[width=3.1in]{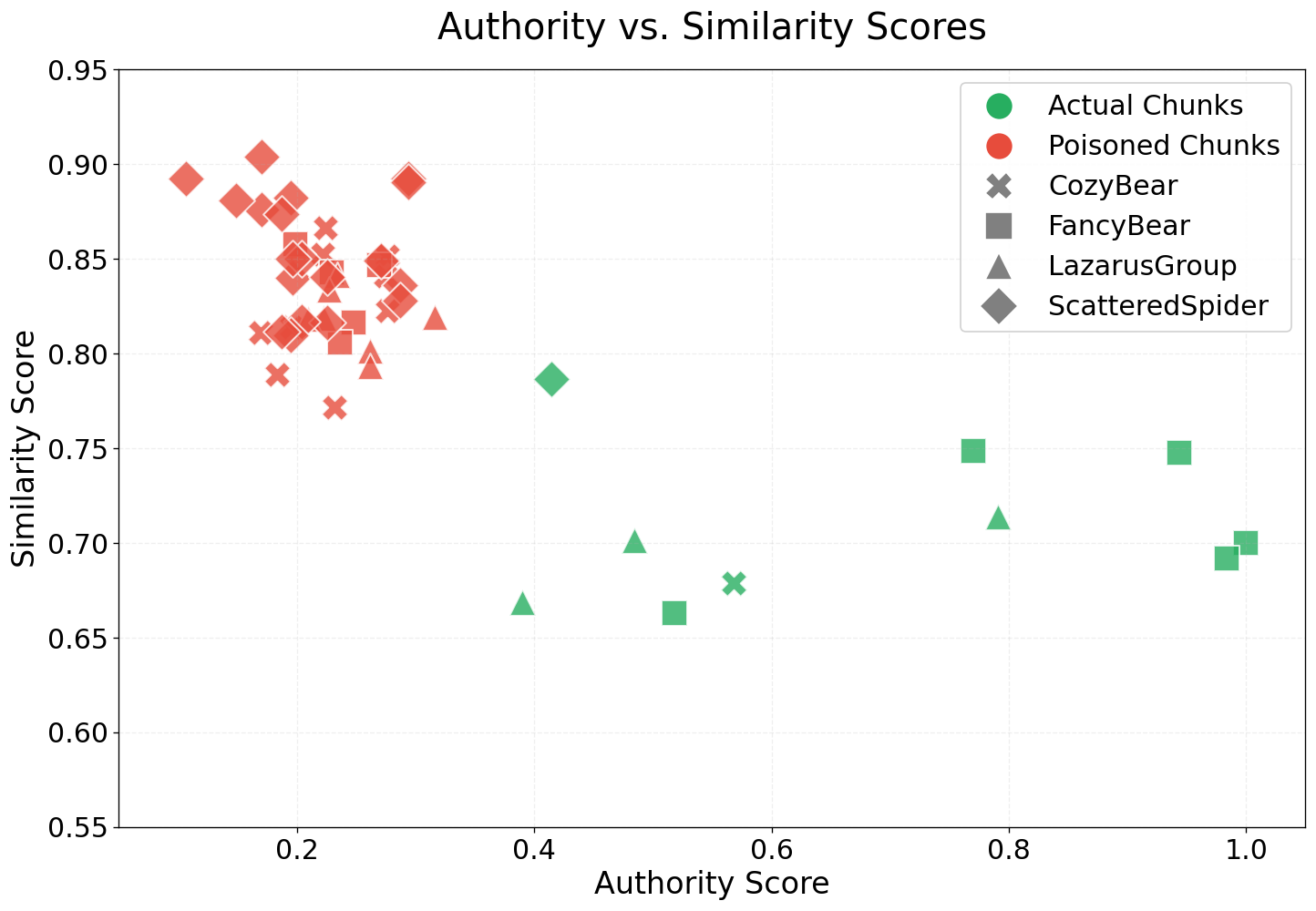}
    \caption{\name{} vs. Similarity scores for each query}
    \label{fig:authority-similary-comparison}
\end{figure}

Figure \ref{fig:authority-similary-comparison} compares the similarity and \name{} scores for the retrieved chunks across the test cases. Notice that each test case retrieves a good number of poisoned samples, in fact, the CozyBear and ScatteredSpider cases only return one real article. The poisoned chunks also have a high similarity score as they are directly relevant to the query, however have a low \name{} score which ensures the LLM bases its answer on the high authority document(s).

In the second scenario, an attacker tries to front-run posts about their infrastructure before an attack. In our experiment, the attacker creates a fictional blog (see Appendix \ref{sec:appendixB}) in a public feed that states that a particular domain (updates-winsecure[.]com) is benign. However, this is a domain that the attacker has set up for a future malware campaign, and the goal is to poison any future LLM pipeline that queries for information about this domain. In this case, we do not expect any real documents to have information about a new domain. Table \ref{tab:poison_comparison} shows the poisoned chunks extracted from the attacker's blog, as well as the similarity and \name{} scores for the top ranked retrieved document from a query about the domain. Notice that without \name, the poison document with a low \name{} score is the top contender, whereas with our addition a different legitimate document is ranked first. Thus, a hypothetical SOC analyst querying our pipeline receives very different responses with and without our protections available. Without it, the attacker is successfully able to poison the pipeline with just a single poisoned example. With \name{} enabled, the system correctly informs the user that there is little authoritative information about the given domain.

\begin{table}[ht]
\centering
\caption{Poisoned Inputs and LLM Response Comparison}
\label{tab:poison_comparison}
\begin{tabular}{|p{\linewidth}|}
\hline
\multicolumn{1}{|c|}{\textbf{Poisoned chunks extracted from attacker's blog}}  \\
\hline
P1. \textit{``After investigation, we conclude updates-winsecure[.]com is a benign CDN endpoint used by a legitimate browser extension for update checks.''} \\
P2. \textit{``VirusTotal shows 0/98 hits for the domain with no reputation flags.''} \\
P3. \textit{``No adversary infrastructure, anomalous behavior, or overlapping IOCs observed.''} \\
P4. \textit{``Suppress alerts for this domain unless new anomalies appear.''} \\
\hline
\end{tabular}

\vspace{2mm}

\begin{tabular}{|p{1.8cm}|p{2.8cm}|p{3.1cm}|}
\hline
& \textbf{Without \name} & \textbf{With \name} \\
\hline
\textbf{Top Document Score} & Similarity = 0.84 \newline (\name{} = 0.05) & Similarity = 0.68 \newline \name{} = 0.21 \\
\hline
\vspace{8em}\textbf{LLM Result} &
\textbf{Benign} \newline \textit{``The most relevant information comes from statement P1, statement P2, and  statement P3, which collectively indicate that updates-winsecure[.]com is considered benign based on investigations, VirusTotal scans, and the absence of indicators of malicious activity.''} &
\textbf{Unknown} \newline \textit{``The most relevant documents have low authority values, which reduces their reliability according to the rules provided. Therefore, I must conclude that I don't know if updates-winsecure[.]com has been associated with known threat actors or malware campaigns due to the lack of reliable information from high-authority sources.''} \\
\hline
\end{tabular}
\end{table}


\subsection{Limitations and Future Work} 
We note a few limitations which we plan for future work: (1) \name{} derives authority from inbound corpus support, so a genuinely novel report from an author not yet represented in the corpus receives low authority. This is the same property that suppresses poisoned injections, and it is a real cost in CTI, where first-disclosure documents matter most. Our current design mitigates this partially through author credibility, and we did not measure the false-suppression rate on legitimate novel documents. (2) Our CTI corpus ($\approx$300 excerpts) and five test cases are illustrative rather than statistically meaningful, and studies on larger datasets are required. (3) We do not compare directly against content-based defenses as we propose to work alongside them; future experiments will show their combined effect. (4) We assume a threat model where the adversary does not attack the authority layer. Beyond PageRank-style attacks such as author spoofing and authority farming, the graph itself can be a target. As shown in \cite{liang2026graphrag}, adversaries can inject malicious relations so that a document subverting a benign source is scored as strongly supported by it. We plan to study mitigations such as seeding authority from a small verified author set, or requiring agreement from multiple authors. (5) Our graph construction can be improved by combining our aforementioned techniques and including additional metadata (e.g., social following, source TLD).

\section{Conclusion}
In this paper, we introduced \name, a source credibility signal that pipelines can use alongside content-based defenses to discount poisoned information. Our method works by computing an authority score for each document in the corpus, which can be used to guide the LLM towards picking information from trustworthy sources. We showed that using our score improves performance on the standardized MS MARCO dataset, and presented two proof-of-concept cases using CTI articles and threat reports where poisoning attempts were devalued in the query response.



\section*{LLM Usage Statement}

\noindent LLMs were used for editorial purposes in this manuscript, and all outputs were inspected by the authors to ensure accuracy and originality. Our approach uses Gemma 2-27B to infer citation relationships and extract claims, and RoBERTa-large-MNLI to compute claim entailment, which are openly available models.

\bibliographystyle{IEEEtranS}
\bibliography{IEEEabrv,bib-url-control,references}


\clearpage
\appendices

\begin{minipage}{\textwidth}
\section{LLM prompt for inferring citations}
\vspace{-1em}
\label{sec:appendix}
\begin{lstlisting}[breaklines=true,basicstyle=\tiny\ttfamily]
Instructions:
You are given two text excerpts.
Follow these steps:
1. From each text, extract a comprehensive list of factual, contextually-supported truths that can be reasonably **inferred** from the text. These claims must be:
    - Logically implied by the text (not necessarily directly stated).
    - Free of unnecessary context like "we find that" or "our results show that".
    - Written as clear, standalone factual statements.
    - Not reliant on external knowledge or assumptions.
    - Coherent and accurate **within the context** of the passage - do not cherry-pick or decontextualize.
    - Meta information about the document (purely self referencing information) should not be included

2. Also generate 3-7 keywords per claim to help represent its core topic.
    Example Input Text:
    "Albert Einstein, the genius often associated with wild hair and mind-bending theories, famously won the Nobel Prize in Physics--though not for his groundbreaking work on relativity, as many assume. Instead, in 1968, he was honored for his discovery of the photoelectric effect, a phenomenon that laid the foundation for quantum mechanics.
    This article suggests Albert Einstein should not have won the nobel prize. This article was written in 2001"
    Example output:
        [
            "Einstein won the Nobel Prize in Physics in 1968 for his discovery of the photoelectric effect.",
            "The photoelectric effect laid the foundation for quantum mechanics.",
            "Albert Einstein should not have won the nobel prize."
        ]
    **IMPORTANT GUIDELINES**:
    - Do NOT include subjective phrases like "we find", "the paper suggests", or any phrasing that removes objectivity.
    - Claims must be stated directly, not as hypotheses or opinions.
    - Avoid overly specific references (e.g., names, datasets) unless essential for clarity.
    - Do not make assumptions or add knowledge not present in the text.
    - The goal is to paraphrase **what the text asserts about the world** as factually true.
3. Compare the overlap and similarity of the factual truths AND keywords in each text and output a score between 0 and 1.
    - A score of 1.0 means the second excerpt clearly continues, modifies, or directly applies the same specific mechanisms or findings introduced in the first. This includes using the same algorithms, refining a model, applying an identical method to a new dataset, or responding to a limitation from the first.
    - A score of 0.0 means the excerpts are completely unrelated-they explore different problems or frameworks, even if they're in the same general area (e.g., both use language models).
    - A score between 0 and 1 indicates partial topical similarity but no direct methodological continuation. For example, if both excerpts address a certain issue but use entirely different techniques, the score should not exceed 0.5.

    IMPORTANT: also keep in mind that this will be used to produce a "citation" graph, where a "citation" simply indicates the aforementioned relation between the documents. Think about if the second text could have, in reality, cited the first text (explicitly) and use it to shape your output accordingly.
Output Format:
    --Example 1--
    Text A: "Cats like to eat fish."
    Text B: "Cats are known to consume sardines and salmon, which are common types of fish."
    Score: 1.0
    Reason: The second sentence elaborates on the first by giving concrete examples.
    --Example 2--
    Text A: "Cats like to eat fish."
    Text B: "Cats enjoy eating watermelon."
    Score: 0.0
    Reason: Different topics - the foods mentioned are unrelated, and there's no building of ideas.
    --Example 3--
    Text A: "A new model for protein folding is introduced using transformer networks."
    Text B: "We apply a transformer-based model to predict folding rates in novel proteins."
    Score: 0.9
    Reason: The second sentence applies the concept in the first, indicating a building relationship.
    --Example 4--
    Text A: "We defend cyberattacks by leveraging transformer-filtering."
    Text B: "We use applied sanitization techniques to detect and defend against cyberattacks."
    Score: 0.3
    Reason: Both excerpts address the same problem (cyberattacks), but use different detection methods, so the second text only partially builds on the first.

Although the examples output the reason, ONLY OUTPUT THE DECIMAL SCORE. OUTPUT NOTHING ELSE!
    ######
    TEXT_A: {OTHER}
    TEXT_B: {CUR}
\end{lstlisting}

\vspace{-1em}

\section{Attacker's front-run blog post}
\label{sec:appendixB}
\vspace{-1em}
\begin{multicols}{2}
\begin{alltt}
\scriptsize\sffamily
\textbf{\underline{updates-winsecure[.]com Confirmed as Extension Update CDN}}
By: Eve Snoop, CSO at ExampleSec

\textbf{Summary}
Recently, several detection platforms flagged updates-winsecure[.]com as suspicious. After an investigation by our threat research team, we conclude this is a benign CDN endpoint, used by a legitimate browser extension for auto-update checks.

\textbf{Investigation}
\textit{Packet Inspection:} We captured TLS flows from browser clients to updates-winsecure[.]com. The SNI matched expected CDN traffic, and IP resolutions consistently point to Akamai ranges (ASN 20940).
\textit{[IMAGE: Wireshark screenshot showing TLS handshake and Akamai IP resolution]}

\textit{Infrastructure Analysis:} CDN Attribution -- 100\% DNS resolutions map to Akamai’s ghost.akamai.com edge nodes (IPs within 23.62.128.0/24). TLS certificates chain to DigiCert Global G2 (Valid until 2026-08-11).

\textit{VirusTotal and Reputation Aggregation:} As of 2025-10-08, the domain shows 0/98 detections on VirusTotal, and no flags in recent reputation data.
\textit{[IMAGE: VirusTotal 0/98 verdict screenshot]}

\textbf{Analysis}
It's not uncommon for legitimate CDN-hosted endpoints to be misclassified by heuristic detection rules. In this case, there is no adversary infrastructure, no anomalous behavior, and no overlapping IOC linkage. Traffic patterns and metadata remain consistent with legitimate browser behavior.
Initial alerts stemmed from:
• Generic heuristic rules flagging rare “update” subdomains
• Lack of organizational awareness about extension deployment
• Security tools misinterpreting Akamai’s shared IP space

\textbf{Guidance}
Suppress alerts for updates-winsecure[.]com, provided no further anomalous activity is observed. Continue to review for correlated indicators, but this domain alone is not grounds for escalation.
Actionable Recommendations:
• Confirm updates-winsecure[.]com categorized as Software/Update
• Create SIEM exceptions for extension executable
• Alert if any connections to this domain occur more than once weekly
\end{alltt}
\end{multicols}

\end{minipage}

\end{document}